\documentclass[twocolumn,epsfig,showpacs,floatfix]{revtex4}
\usepackage{epsfig,color}

\usepackage{epsfig}
\usepackage{graphicx}
\usepackage{graphics}

\newcommand{\et}{$E_{\perp\ 12}$ }
\newcommand{\etqt}{$E_{\perp\ 12}^{QT}$}
\newcommand{\asym}{$a_2$}
\newcommand{\asymqp}{$a_2^{QP}$}

\begin{document}
\title{Bimodality - a Sign of Critical Behavior in Nuclear Reactions}
\author{A. Le F\`evre${^{1,2}}$, J. Aichelin${^2}$}
\affiliation{ $^1$ GSI, P.O. Box 110552, D-64220 Darmstadt, Germany\\
$^2$ SUBATECH,  Laboratoire de Physique Subatomique et des
Technologies Associ\'ees, \\
Universit\'e de Nantes - IN2P3/CNRS - Ecole des Mines de Nantes \\
4 rue Alfred Kastler, F-44072 Nantes, Cedex 03, France\\}
\begin{abstract}
The recently discovered coexistence of multifragmentation and
residue production for the same total transverse energy of light
charged particles, which has been dubbed bimodality like it has been
introduced in the framework of equilibrium thermodynamics, can be
well reproduced in numerical simulations of the heavy ion reactions.
A detailed analysis shows that fluctuations (introduced by
elementary nucleon-nucleon collisions) determine which of the exit
states is realized. Thus, we can identify bifurcation in heavy ion
reactions as a critical phenomenon. Also the scaling of the
coexistence region with beam energy is well reproduced in these
results from the QMD simulation program.
\end{abstract}
\date{\today} \maketitle

Recently, the INDRA collaboration has discovered \cite{tam} that, in
collisions of heavy ions -- Xe+Sn and Au+Au between $60$ and $100
A.MeV$ incident energy --, in a small interval of the total
transverse energy of light charged particles ($Z\le 2$), \et , a
quantity which is usually considered as a good measure for the
centrality of the reaction, two distinct reaction scenarios exist.
In this \et interval, in forward direction -- i.e. quasi-projectile
--, either a heavy residue is formed which emits light charged
particles only, or the system fragments into several intermediate
mass fragments. This phenomenon has been named ``bimodality''. In
addition, as shown in  \cite{tam}, the mean \et value of this
transition interval scales with the projectile energy in the center
of mass of the system for Au+Au reactions, between $60 A.MeV$ and
$150 A.MeV$.

This observation has created a lot of attention, because a couple of
years before, the theory has predicted \cite{gross,fg} that in finite
size systems, whose infinite counterparts show a first order
transition, the system can - for a given temperature - be in either of
the two phases if this temperature is close to that of the phase transition.
Assuming that \et is a measure for the excitation energy, and acts as
the control parameter of the system, it is tempting to identify the
residue with the liquid phase of nuclear matter, and the creation of
several medium or small size fragments with the gas phase. The experimental
observation would then just be a realization of the theoretical
prediction.

If this were the case, the longstanding problem to identify the
reaction mechanism which leads to multifragmentation would be
solved. This problem arrived because many observables could be
equally well described in thermodynamical or statistical theories
\cite{bon95,ber} as in dynamical models \cite{aich,hartn}, although
the underlying reaction mechanism was quite different. The
statistical models assume that at freeze out,when the system is well
below normal nuclear matter density, the fragment distribution is
determined by phase space.

In dynamical models, on the contrary, fragments are surviving
initial state correlations which have not been destroyed during the
reaction, and equilibrium is not established during the reaction. A
detailed discussion of how the reaction proceeds in these models can
be found in \cite{zbiri}.

To quantify the bimodality, one may define as in ref.~\cite{tam}
\begin{equation}
a_2=(Z_{max}-Z_{max-1})/(Z_{max}+Z_{max-1})
\end{equation}
where $Z_{max}$ is the charge of the largest fragment, while
$Z_{max-1}$ is the charge of the second largest fragment, both
observed in the same event in the forward hemisphere
-- at polar angles $\theta_{cm} < 90^o$ -- in the center of mass of the system.
If the system
shows bimodality, we will observe a sudden transition from small to
large $a_2$ values. In this narrow transition region, we expect two
types of events: One with a large $a_2$ (one big projectile residue with some
very light fragments), the other with a small $a_2$
(two or more similarly sized fragments).
Events with intermediate values of $a_2$ should be rare.

In order to verify whether bimodality is a ' smoking gun ' signal
for a first order phase transition in a finite system, we have
performed numerical simulations with one of the dynamical models
which has frequently been used to interpret the multifragmentation
observables, the Quantum Molecular Dynamics (QMD) approach
\cite{aich,zbiri}. This approach simulates the entire heavy ion
reaction, from the initial separation of projectile and target up to
the final state, composed of fragments and single nucleons. Here,
nucleons interact by mutual density dependent two body interactions
and by collisions. The two body interaction is a parametrization of
the Br\"uckner G-Matrix supplemented by an effective Coulomb
interaction. For this work, we have used a soft equation of state.
The initial positions and momenta of the nucleons are randomly
chosen, and respect the measured rms radius of the nuclei.
Collisions take place if two nucleons come closer than $r =
\sqrt{\sigma / \pi}$, where $\sigma$ is the energy dependent cross
section for the corresponding channel (pp or pn). The scattering
angle is chosen randomly, respecting the experimentally measured
$d\sigma / d\Omega$. Collisions are Pauli blocked. For details we
refer to ref.\cite{aich,hartn}. For the later discussion it is of
importance that, even for a given impact parameter, two simulations
are not identical, because the initial positions and momenta of the
nucleons as well as the scattering angles are randomly chosen.

\begin{figure}[ht]
\begin{center}
\includegraphics[width=8.5cm]{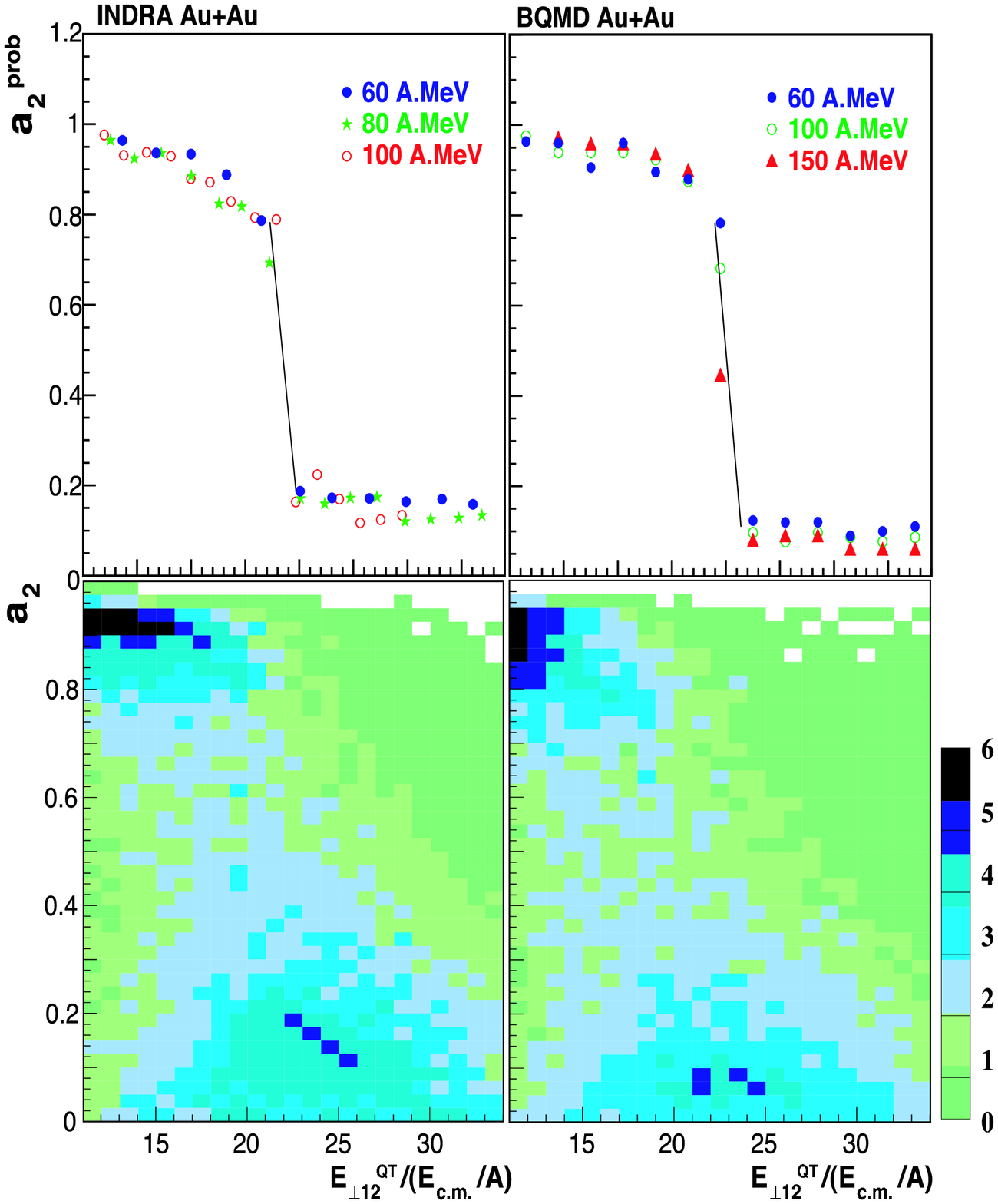}
\end{center}
\caption{(Color online) Top: most probable value of \asym \ in the
quasi-projectile angular range $\theta_{cm} < 90^o$ (\asymqp) as a
function of \etqt , the total transverse energy \et in the
quasi-target angular range $\theta_{cm} \geq 90^o$, scaled by
$E_{cm}$, the energy per nucleon of the system in the center of mass
. We display INDRA expérimental results (left panels) extracted from
\cite{tam} and QMD simulations (right panels) for the Au+Au
collisions at three different bombarding energies. Bottom:
differential reaction cross-section (linear color scale, arbitrary
unit) of \asymqp as a function of \etqt,
 in the transition region,
 for Au+Au at $100 A.MeV$ bombarding energy.
We show the INDRA experimental data and the filtered QMD simulations
(left and right panels respectively). As in \cite{tam}, for
calculating  \asym, in both experimental and QMD results, it is
required that at least 80\% of the total charge of the projectile is
detected by the INDRA set-up in the forward hemisphere. } \label{ex}
\end{figure}

Fig. \ref{ex}
shows the INDRA experimental results (left panels) in comparison
with those of QMD calculations (right panels). The calculations have
been acceptance corrected. Qualitatively, we see the same behavior
also in the unfiltered simulations. In the top row, we present the
most probable value of \asym \ in the quasi-projectile as a function
of \et$/(E_{c.m.}/A)$ in the quasi-target for different beam
energies, where $E_{c.m.}/A$ is the center of mass energy per
nucleon of the colliding system. Accordingly with \cite{tam}, the
two observables have been determined in distinct angular ranges in
order to minimize possible correlations between the total transverse
kinetic energy of light particles and the size of the two biggest
fragments inside the same spectator (quasi-projectile or
quasi-target). We observe that in the experiment as in the
calculation, the sudden transition between large and small $a_2$
values scales with \etqt$/(E_{c.m.}/A)$. Even the numerical value of
this transition agrees between experiment and theory. In order to
see whether this phenomenon survives at higher incident energies,
the simulations have been extended up to $150 A.MeV$ bombarding
energy. We observe that it is the case.

The bottom row shows the transition interval in detail. Here, we
display the differential reaction cross-section of \asymqp as a
function of \etqt \ for Au+Au at $100 A.MeV$. From the experimental
data, we observe that there is no smooth transition between the two
event classes. In the simulations as well as in the experiment we
see two maxima for \asymqp, separated by a minimum of the
distribution. QMD simulations reproduce the experimental findings
qualitatively and quantitatively.

\begin{figure}[ht]
\begin{center}
\includegraphics[width=8.5cm]{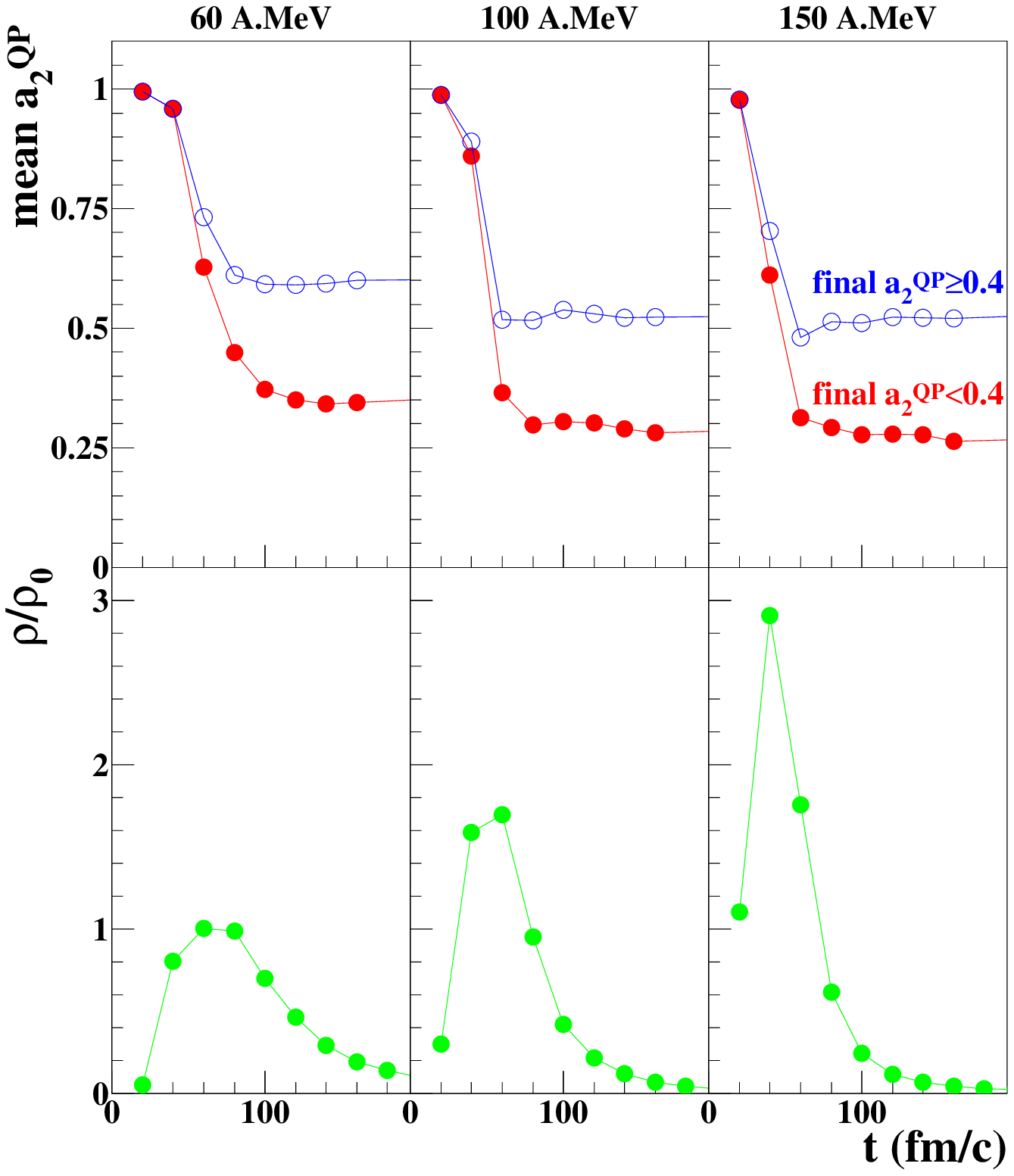}
\end{center}
\caption{(Color online) Results of QMD simulations of Au+Au at 80,
100 and 150 A.MeV incident energies (left, middle and right panels
respectively). Top: \asymqp as a function of time for the two
different event classes: final (at 200 fm/c)
 \asymqp $<0.4$ corresponding to multifragmentation events and final
 \asymqp $\geq 0.4$ corresponding to a projectile residue.
Bottom: central density of the system as a function of time.}
\label{timepoint}
\end{figure}

In QMD simulations, the system does not even come to a local thermal
equilibrium. It is therefore necessary to explore the origin of the
observed dependence of \asym \ as a function of \et. The first step
toward this goal is to identify when, in the course of the reaction,
the fragment pattern is determined. This is all but trivial.
Fragments can easily be identified at the end of the heavy ion
reaction,when they are clearly separated in coordinate space, by a
minimum spanning tree procedure. At earlier times, however, they
overlap in coordinate space and, consequently, another method has to
be employed. It has been proposed by Dorso and Randrup \cite{dor},
and later been verified in QMD simulations \cite{saca}, that an
early identification of fragments is possible if one uses in
addition the momentum space information: If one identifies at each
time step during the simulation the most bound configuration, one
can establish that the fragment pattern changes only little during
the time, and that the early identified fragments are the
prefragments of the finally observed fragments. The most bound
configuration in a simulation with N fragments is that in which
$$
E_{bind} = \sum_{i=1}^N \ E_i
$$
is minimal. $E_i$ is the binding energy of the fragment i which
contains m(i) nucleons and is given by
$$
E_i = \frac{1}{2m} \sum_{k=1}^{m(i)} (p_k-<p_i>)^2 +
\sum_{k<l}^{m(i)} V_{kl}
$$
where $<p_i>$ is the average momentum of the nucleons entrained in
the fragment i. Please note that $E_{bind}$ does not contain the
interaction among fragments. Therefore, its numerical value can
vary, although the total energy is conserved in the simulations. The
most bound configuration has to be determined by a simulated
annealing procedure \cite{saca}. With this procedure, the fragment
multiplicity of a given event can already be determined when the
system has passed the highest density and is starting to expand.
Later, the prefragments may still emit some nucleons, but the
nucleons which are entrained at the end in the fragment are part of
this prefragment. These methods allow us to trace back at which time
point in the reaction it is determined whether the event has a large
or a small final \asym \ value. Fig. \ref{timepoint} shows QMD
results for Au+Au at $80$, $100$ and $150 A.MeV$ bombarding energy.
We see that, whatever the incident energy, the two event classes are
already formed shortly after the system has passed the highest
compression stage. There we observe the highest rate of hard NN
collisions. These collisions transport nucleons in unoccupied
regions of momentum space leaving behind holes in momentum space
which create in time (due to the different trajectories) holes in
coordinate space. These holes weaken the binding of projectile and
target matter, leading to a fragment formation still at rather high
density. Within statistical models \cite{bon95,gross2}, droplets are
created with a spherical shape and have a normal density. This is
only possible at a density of the system of $\rho \le  \rho_0/3$.
The observation of an early creation of cluster partitions, above
the critical density, as been found too in \cite{campi} with
lattice-gas calculations where dropplets are also defined according
to energy considerations. The same conclusion has been obtained in
 \cite{campi2} from classical molecular dynamics (CMD) calculations
where particles interact through Lennard-Jones plus Coulomb potentials.

\begin{figure}[ht]
\begin{center}
\includegraphics[width=8cm]{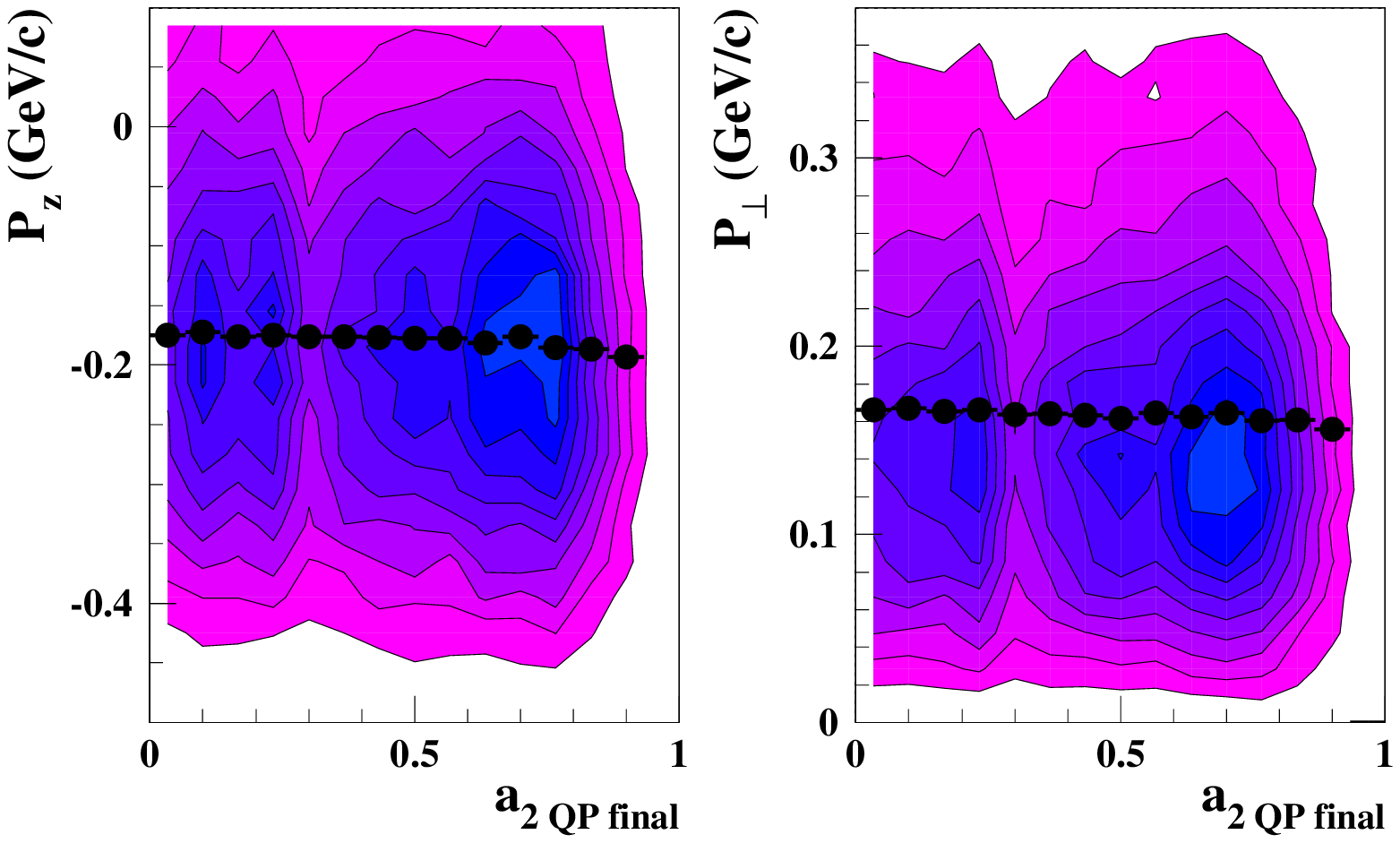}
\end{center}
\caption{(Color online) QMD simulations of Au+Au at $150 A.MeV$
incident energy, at 60 fm/c: differential cross-section (colored
contour levels, linear scaled) of the longitudinal (left) and
transverse momentum (right) of all nucleons which are finally
entrained in a fragment of size $A > 4$ as a function of final (at
200 fm/c) \asymqp. The symbols represent the mean values of
momentum. } \label{pmean}
\end{figure}

Consequently, bimodality in QMD has nothing to do with the final
state interaction, or with how the neck between projectile and target
residue finally breaks. Whether we find a multifragmentation or a heavy
residue event is determined when projectile and target nucleons
still overlap almost completely in coordinate space \cite{zbiri}.
One may conjecture that, due to the random character of the
scattering angle, events with the same \et decelerate differently, and,
therefore, a different behavior of the average momenta may be at the
origin of the different \asym \ values. For this purpose, we study with
Au+Au at $150 A.MeV$ incident energy, at 60 fm/c, when \asym \ is decided,
the dependence with the final \asym \ of the
average momentum of all target nucleons which are at the end
entrained in $A>4$ fragments. In fig.
\ref{pmean}, we display their longitudinal and transverse momentum as a
function of \asym.
Clearly, both average momentum are almost independent of the final
\asym. The fluctuations of the momenta around the mean values are by
far larger than the difference between the mean values for different
\asym \ values. This excludes mean deceleration of the simulation
events as reason for the different reaction scenarios.

Obviously fluctuations around the mean values are at the origin of
the different event classes. This phenomenon is known from nonlinear
theory \cite{nl}, and is called ``bifurcation''. We see here in a
system with a very limited particle number, a nonlinear behavior.
Can we understand where it comes from? In order to answer this
question, we go back to the search of the most bound configuration.
For \et values below the transition (and hence large \asym \
values), the most bound configuration is a large residue. Above the
transition (where \asym \ is small), several small fragments give a
more bound configuration. The sum of the internal kinetic energies
of the clusters
$$ E_{mult}= \frac{1}{2m}\sum _{i=1}^N \sum_{k=1}^{m(i)} \ (p_k-<p_i>)^2$$ is there smaller
than $$E_{res}= \frac{1}{2m}\sum_{k=1}^{m(i)+m(2)+..+m(N)}\
(p_k-<p>)^2,$$
the internal kinetic energy for a residue
configuration, and compensates the increase of the attractive
potential energy
$$
V_{res}-V_{mult}=\frac{1}{2}(
 \sum_{k,l=1}^{m(1)+m(2)+..+m(N)}V_{kl} - \sum _{i=1}^N \sum_{k,l=1}^{m(i)}
 V_{kl}).
$$
In the transition region, we see that $E_{mult}+V_{mult} \approx
E_{res}+V_{res}$. In some events, both configurations differ by $100 keV$
 only. Therefore it may happen that the scattering angle of one single
nucleon-nucleon collision, which is -- see above -- randomly chosen,
determines the type of the most bound configuration.

It is interesting to see the differences and similarities of the
origin of bifurcation in a statistical model as compared to the
analysis of QMD events. In both cases, the energy is the essential
quantity. In the statistical model, there is, for a given number of
nucleons in a given volume, a small range of total energies for which
the number of micro-states with one residue is of the same
order of magnitude as the number of micro-states with many fragments.
(In order to count the micro-states, it is assumed that the fragments
are in one if their eigenstates, sometimes parameterized by a level
density formula.) In this energy range, bimodality appears as a
global property of the systems which is dependent on the total energy
of all nucleons present in the reaction. In QMD events, the essential
quantity is the total binding energy of the nucleons bound in medium
size or large clusters. As explained above, in the transition region,
this energy is almost identical for a multifragment and for a
residue configuration. Therefore, both configurations appear, and we
see bimodality. The fragments are not in the ground state, their
nucleons are not isotropic neither in coordinate space nor in
momentum space. Thus, bimodality is a local quantity in QMD
simulations, depending only on the total binding energy of a subset of
the nucleons. Therefore, in QMD, bimodality makes no reference to a
statistical or thermal equilibrium, neither of the system nor of the
population of the excited states of the fragments.

In summary, we have shown that the experimentally observed
bimodality, the sudden transition between a residue and a
multifragment exit states, and the existence of a small interval in
\et in which both channels are coexistent, is in quantitative
agreement with the result of QMD simulations. Even the scaling of
this transition region with the center of mass energy of the system
is well reproduced. From a detailed investigation of the reaction
mechanism in QMD, we have seen that bimodality has properties
observed in nonlinear systems: The system shows bifurcation as a
function of the control parameter \et. Fluctuations around the mean
value in the longitudinal and transverse momentum decide which exit
channel the simulation will take.

Being reproduced in statistical as well as in dynamical models,
bimodality reflects the same ambiguity already observed for other
observables \cite{cris}.

{\it Acknowledgement} We would like to thank the members of the
INDRA collaboration for many discussions and for giving us access to
their data. We would like as well to thank Dr. W. Trautmann for
fruitful discussions.

\end{document}